\documentclass[%
pra,
superscriptaddress,
 tightenlines,
showpacs,showkeys,
12pt,
a4paper
]{revtex4}
\pdfoutput=1
\usepackage{amssymb,amsmath,stmaryrd,array}

\usepackage{graphicx}
\usepackage{subfig}

\makeatletter

\newcommand{\cnj}[1]{{#1}^{\ast}}

\newcommand{\tcnj}[1]{{#1}^{T}}

\newcommand{\pdrs}[1]{\partial_{#1}}

\newcommand{\mum}{$\mu$m}

 \newcommand{\vc}[1]{\mathbf{#1}}
 \newcommand{\mvc}[1]{\mathbf{#1}}
 \newcommand{\uvc}[1]{\hat{\mathbf{#1}}}

\newcommand{\dd}{\mathrm{d}}

\newcommand{\inc}{\mathrm{inc}}

\newcommand{\trans}{\mathrm{tr}}
\newcommand{\vac}{\mathrm{vac}}
\newcommand{\med}{\mathrm{m}}

\makeatother

\begin{document}
\DeclareGraphicsExtensions{.eps,.png,.pdf}
\title{
Conoscopic patterns in photonic band gap of cholesteric liquid crystal cells  
with twist defects
}

\author{Roman~I.~Egorov}
\email[Email address: ]{rommel@iop.kiev.ua}

\affiliation{%
 Institute of Physics of National Academy of Sciences of Ukraine,
 prospekt Nauki 46,
 03028 Ky\"{\i}v, Ukraine}

\author{Alexei~D.~Kiselev}
\email[Email address: ]{kiselev@iop.kiev.ua}

\affiliation{%
 Institute of Physics of National Academy of Sciences of Ukraine,
 prospekt Nauki 46,
 03028 Ky\"{\i}v, Ukraine}

\date{\today}

\begin{abstract}
We theoretically investigate into
the effects of the incidence angles  in 
light transmission of
cholesteric liquid crystal
two-layer sandwich structures with 
a twist defect created by rotation of the one layer 
about the helical axis. %by the defect twist angle.
The conoscopic images and polarization resolved patterns
are obtained for thick layers
by computing the intensity and the polarization parameters 
as a function of the incidence angles.
In addition to the defect angle induced rotation
of the pictures as a whole,
the rings of defect mode resonances
are found to shrink to the central point
and disappear
as the defect twist angle varies from zero to 
its limiting value $\pi/2$ and beyond.
\end{abstract}

 \maketitle

%%%%%%%%%%%%%%
\section{Introduction}
\label{sec:intro}
%%%%%%%%%%%%%%

Equilibrium orientational structures in cholesteric liquid crystals (CLC)
are represented by helical twisting patterns
where molecules on average rotate in a helical fashion  about a  uniform twist axis.
These chirality induced helical structures are responsible for
distinctive optical properties of CLCs that  have been
the subject of  intense studies over the last few decades.

It is well known that,
similar to one dimensional (1D) photonic crystals,
the ideal CLC helix (the planar Grandjean structure)
is characterized by a photonic band gap
(an optical stop band)
where the circular polarized light
with the  helicity identical to the handedness of the helix
cannot propagate and the selective 
reflection takes place~\cite{Belyakov:bk:1989}.

In the most studied case of normal incidence,
where the light is propagating along the twist axis
(the $z$ axis) of the CLC helix, which is characterized by 
the director field, $\uvc{d}=(\cos(2\pi z/P+\phi_0),\sin(2\pi z/P+\phi_0),0)$,
(the unit vector giving the direction of preferred orientation of CLC molecules)
and the CLC pitch, $P$,
the Maxwell equations can be solved 
analytically~\cite{Belyakov:bk:1992,Good:josa:1994}.
The well-known result is that
the wavelength of light $\lambda$ falls between
$n_o P$ and $n_e P$ in the stop band,
where $n_o$ and $n_e$ are the ordinary and extraordinary
indices, respectively. 

Similar to the case of isotropic photonic
materials~\cite{Joannop:bk:2002}, 
when
the one dimensional (1D) periodicity of
anisotropic helical structures in CLCs  
is disrupted,
there are
a number of physical effects
arising due to the presence of localized modes.
For instance, in the reflection/transmittance spectral curves of CLCs, 
the localized modes manifest themselves as 
peaks and dips that take place for both circular polarizations of
incident light and 
located within the selective reflection band gap.

These modes can be produced within the photonic band gap 
by a variety of defects such as
isotropic defect layers~\cite{Yang:pre:1999,Belyakov:mclc:2008},
twist defects~\cite{Kopp:prl:2002,Kopp:optl:2003,Wang:prsa:2005,Chen:jpd:2005},
%spatially varying pitch combined with twist defects\cite{Chen:jpd:2005}
variations of the CLC pitch~\cite{Stille:epje:2003,Ozaki:jap:2006}
and local deformations of the CLC helix~\cite{Matsui:pre:2004}.

Defect modes in CLCs have received much attention
because they are  known to play the key part in
photonic devices such as low-threshold lasers~\cite{Kopp:prqe:2003,Fuh:optex:2004,Song:advmat:2006,Palto:jetp:2006}
and polarization dependent filters~\cite{Chen:jopa:2005}.
But, in the bulk of studies,
a major focus of interest is
the case of normal incidence where the waves propagate along 
the helical axis,
whereas effects due to oblique incidence 
(off-axis propagation) have not yet been adequately explored.
In the very recent paper~\cite{Arkhipkin:jetp:2008}, 
it was found that
the defect modes in 1D isotropic photonic crystal with a defect layer 
represented by a planar oriented nematic liquid crystal film
are very sensitive to the variations of the angle of incidence.

\begin{figure*}[!tbh]
\centering
   \resizebox{70mm}{!}{\includegraphics*{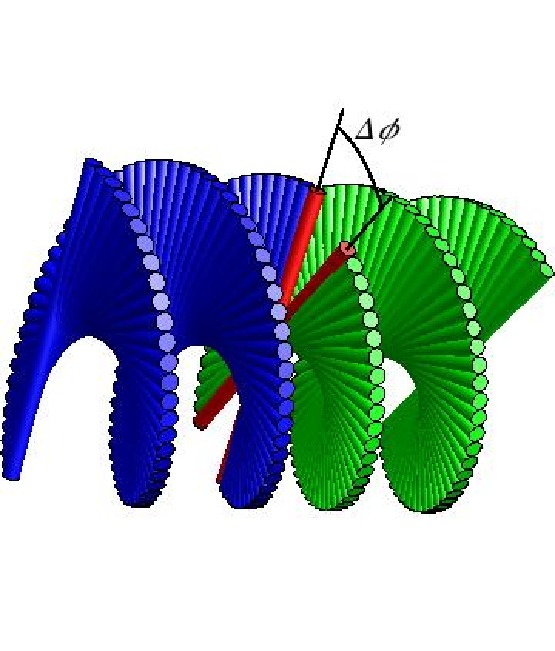}}
\caption{%
Twist defect between two cholesteric helices
with the defect twist angle $\Delta\phi$. 
}
\label{fig:defect}
\end{figure*}

In this Letter we intend to fill the gap and theoretically examine 
the optical properties of the CLC two-layer sandwich structure with 
a twist defect created by rotation of the one layer 
about the helical axis by the defect angle $\Delta \phi$ 
(see Fig.~\ref{fig:defect}).
The characteristics of the transmitted light within the photonic band gap 
at varying incidence angles and polarization of incident wave 
will be of our primary interest.
In particular, we present the results on 
both the standard conoscopic images (intensity patterns)~\cite{Bjorknas:lc:2003,Song:josa:2008} 
and the polarization-resolved angular patterns~\cite{Kis:jpcm:2007,Kiselev:pra:2008}.

\begin{figure*}[!tbh]
\centering
\subfloat[]{
  \resizebox{90mm}{!}{\includegraphics*{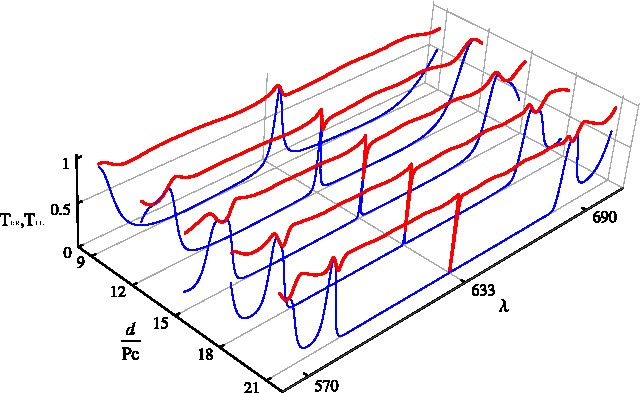}}
\label{subfig:crossover}
}
\subfloat[]{
  \resizebox{70mm}{!}{\includegraphics*{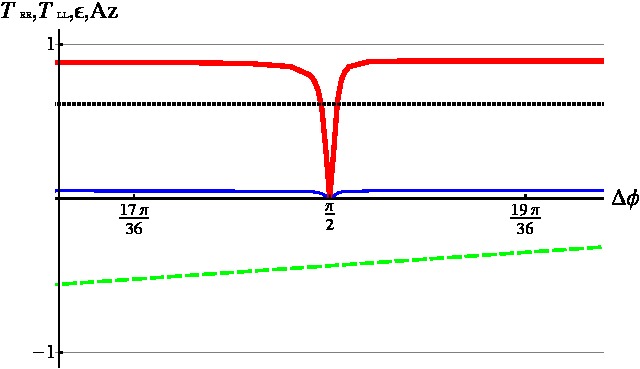}}
\label{subfig:transm}
}
\caption{%
(a)~Spectral curves for the co-polarized transmission amplitudes
$T_{LL}$ (thick red lines) and $T_{RR}$ (thin blue lines)
of light normally incident on
the CLC layer with the twist defect at $\Delta\phi=\pi/2$
computed at varying cell thickness, $D$.  
(b)~Transmission amplitudes as a function of the twist angle,
$\Delta\phi$, at $P=\lambda/n_{c}$ and $D=25$~\mum.
Black dotted and green dashed lines represent 
the ellipticity, $\epsilon$,
and the polarization azimuth, $Az$, respectively. 
}
\label{fig:cross_transm}
\end{figure*}

%%%%%%%%%%%%%%
\section{CLC layer with twist defect}
\label{sec:simulation}
%%%%%%%%%%%%%%

In our calculations, we assume that
the surrounding medium is isotropic and its refractive
index $n_{\med}$ is unity (air),
whereas the refractive indices of CLC,
$n_{o}=1.52$ and $n_{e}=1.71$,
are typical of chirally doped liquid crystal mixtures E7
(commercially available from Merck).
The value of the CLC pitch, $P=\lambda_c/n_b$, 
where $n_b=(n_o+n_e)/2$, is taken to be at the centre of 
the photonic band gap for normally incident irradiation with a He-Ne
laser of the wavelength $\lambda_c=633$~nm.

We also restrict ourselves to the symmetric case where 
the CLC layers are identical in thickness, $D_1=D_2\equiv D$.
According to Refs.~\cite{Kopp:prl:2002,Wang:prsa:2005},
when the thickness changes, 
the defect modes resonances in
the transmission spectra of normally incident light
transform giving rise to the crossover 
phenomenon related to 
 the two limiting structures: 
peaks and dips that occur in the regimes of thin and thick layers, respectively.
This phenomenon is illustrated in Fig.~\ref{subfig:crossover}
which shows the spectral curves for the diagonal elements of the
transmission matrix, $T_{RR}=|t_{--}|^2$ and $T_{LL}=|t_{++}|^2$,
\begin{align}
\label{eq:transm_matrix}
\mvc{T}=
\begin{pmatrix}
  t_{++}& t_{+-}\\
  t_{-+}& t_{--}
\end{pmatrix}
=
\mvc{T}_2\cdot
\left[
\mvc{I}_2 - \tcnj{\mvc{R}}_1\cdot\mvc{R}_2
\right]^{-1}\cdot \mvc{T}_1,
\end{align}
where $\mvc{I}_n$ is the $n\times n$ identity matrix; 
the superscript $T$ indicates the matrix transposition;
$\mvc{T}_i$ ($\mvc{R}_i$) is 
the transmission (reflection) matrix of 
the $i$th layer, 
linking the circular components of the incident and transmitted light,
$E_{\pm}^{(\inc,\,\trans)}=(E_{\parallel}^{(\inc,\,\trans)}\mp iE_{\perp}^{(\inc,\,\trans)})/\sqrt{2}$.

Referring to Fig.~\ref{subfig:crossover},
the co-polarized transmission amplitude for right-handed circular polarized light, 
$T_{RR}$, undergoes the peak-to-dip transition as the thickness
increases reaching the regime of thick cells,
where the defect mode manifests itself as the pronounced dip
in spectral curves for the coefficient $T_{LL}$. 
Since this regime is less studied as compared to the case of
thin cells, in this work, it will be of our primary concern. 
So,  we deal with sufficiently thick CLC layers of the thickness $D=25$~$\mu$m. 

Our computational procedure 
is based on the theoretical approach developed in 
Refs.~\cite{Kis:jpcm:2007,Kiselev:pra:2008}.
In this method, the transfer and reflection matrices,
$\mvc{T}$ and $\mvc{R}$,
are expressed in terms of the evolution operator
of the Maxwell's equations rewritten in 
the $4\times 4$ matrix form
giving the system for 
the in-plane components
of the electric and magnetic fields
combined into
the vector
$\mvc{F}=\tcnj{\bigl(E_x(z),E_y(z),H_y(z),-H_x(z)\bigr)}$:
$-i\pdrs{\tau}
 \vc{F}
 =\mvc{M}\cdot\vc{F}$,
where $\tau\equiv k_{\vac} z=2\pi z/\lambda$ and  
$\pdrs{\tau}$ stands for the derivative with respect to $\tau$.
So, the first step involves computing the evolution operator, $\mvc{U}$,
as the solution to the initial-value problem
\begin{align}
\label{eq:system}
 -i\pdrs{\tau}
 \mvc{U}(\tau|\tau_0)
 =\mvc{M}(\tau)\cdot\mvc{U}(\tau|\tau_0),
\quad
\mvc{U}(\tau_0|\tau_0)=I_4,
\end{align}
where $\mvc{M}$ is the matrix determined by
the CLC dielectric tensor, the incidence
angle, $\theta_{\inc}$, and 
the azimuthal angle of the incidence plane,
$\phi_{\inc}$
(the general expression for $\mvc{M}$ 
can be found, e.g., in~\cite{Kis:jpcm:2007,Kiselev:pra:2008}). 
 
For normally incident light with $\theta_{\inc}=0$,
it can be shown that
the rotating wave ansatz makes the matrix of the
system~\eqref{eq:system} independent of $\tau$
and thus the evolution operator can be found in the closed form.
The transmission matrix~\eqref{eq:transm_matrix}
then can be readily calculated by using the relation
for the $2\times 2$ block structure of the linking matrix~\cite{Kis:jpcm:2007,Kiselev:pra:2008}
\begin{align}
 \label{eq:W-op}
  \mvc{W}=
\mvc{V}_{\med}^{-1}\cdot\mvc{U}^{-1}(h|0)\cdot\mvc{V}_{\med}=
\begin{pmatrix}
\mvc{T}^{-1} & \mvc{W}_{12}\\
\mvc{R}\cdot \mvc{T}^{-1} & \mvc{W}_{22}
\end{pmatrix},
\end{align} 
 where $h=k_{\vac} (D_1+D_2)$ and 
$\mvc{V}_{\med}$ is the matrix of eigenvectors for
the ambient medium.
In Fig.~\ref{subfig:crossover},
we used the analytical results to plot the transmission amplitudes,
$T_{RR}$ and $T_{LL}$, in relation to the wavelength, $\lambda$.
Dependence of both amplitudes on the defect twist angle
$\Delta\phi$ depicted in Fig.~\ref{subfig:transm}
demonstrates a pronounced dip at $\Delta\phi=\pi/2$.
This figure additionally shows
the ellipticity, $\epsilon$, and polarization azimuth,
$Az$, of the transmitted wave computed for the left-handed circular
polarized incident light, so that
$\epsilon=(|t_{++}|-|t_{-+}|)/(|t_{++}|+|t_{-+}|)$
and $Az=\arg{(\cnj{t}_{++}t_{-+})}/2$.

As demonstrated in Fig.~\ref{subfig:transm}, in thick CLC cells, 
the ellipticity remains unchanged, whereas the azimuth  varies
linearly with the defect angle: $Az(\Delta\phi)-Az(0)=\Delta\phi$.
This behaviour results from the fact that, 
in this regime, the polarization characteristics of the transmitted light
are independent of the incident wave polarization.
From the analytical expressions~\eqref{eq:transm_matrix}, it can also be inferred that
orientation of the polarization ellipse is determined by the
azimuthal angle of the CLC director at the exit face, $z=D_1+D_2$.

\begin{figure*}[!tbh]
  \centering
   \resizebox{170mm}{!}{\includegraphics*{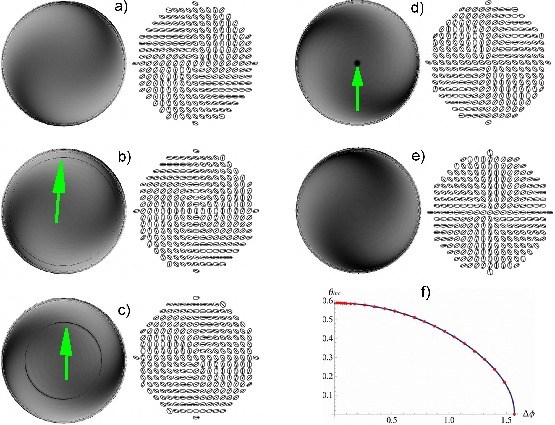}}
\caption{%
(a)-(e)~Intensity conoscopic patterns 
for the co-polarized transmission amplitude
$T_{LL}$
(left) and  
distributions of polarization ellipses 
for LHCP incident light (right)
at different defect angles:
(a)~$\Delta\phi=0$~deg;
(b)~$\Delta\phi=30$~deg;
(c)~$\Delta\phi=60$~deg;
(d)~$\Delta\phi=90$~deg;
(e)~$\Delta\phi=135$~deg.
Arrows point to the rings of defect mode resonances.
(f)~Incidence angle of defect mode resonance ring as a function of
defect angle.
}
\label{fig:pattern}
\end{figure*}

%%%%%%%%%%%%%%%%%%%%%%%
\section{Results and discussion}
\label{sec:results}
%%%%%%%%%%%%%%%%%%%%%%

The case of oblique incidence (off-axis propagation)
with $\theta_{\inc}\ne 0$
is not exactly solvable and the system~\eqref{eq:system}
has to be solved numerically.
In Fig.~\ref{fig:pattern}, we present the results of calculations 
in the form of the traditional intensity conoscopic patterns
representing the angular distributions of the co-polarized transmission amplitude
$T_{LL}$ in the plane of observation where the polar coordinates,
$(\rho,\phi)$ are determined by the incidence angles:
$\rho\propto \tan\theta_{\inc}$ and 
$\phi=\phi_{\inc}$~\cite{Bjorknas:lc:2003,Song:josa:2008,Kiselev:pra:2008}.
Note that the dominating contribution
to the transmittance comes from this amplitude
because, as is evident from Fig.~\ref{fig:cross_transm}, 
$T_{LL}\gg T_{RR}$.  

The conoscopic patterns for the 
defectless sandwich  structure
 with $\Delta\phi=0$ representing the CLC layer of 
the thickness  $D_1+D_2=2D$
are shown in Fig.~\ref{fig:pattern}a.
It is well known that
the selective reflection region shifts to
shorter wavelengths (blue shift)
as the incidence angle $\theta_{\inc}$ increases~\cite{Bjorknas:lc:2003}.
In our case where $\lambda_{c}=633$~nm and 
$\Delta\phi=\pi/2$, 
the points are found to fall outside the photonic band gap
provided the incidence angle is above $40$~deg.
So, the patterns are computed for the solid angle with
$\theta_{\inc}\le 35$~deg so as to have the images
for the distributions within the photonic band gap.   

Without the defect
there are several known features~\cite{Bjorknas:lc:2003}  
such as Airy spirals 
that are visible in Fig.~\ref{fig:pattern}a.
In the corresponding polarization-resolved pattern depicted as the field of
polarization ellipses, the ellipticity varies smoothly
and its value ranges from $0.6$ to $0.8$.
By contrast, the polarization azimuth undergoes noticeable
changes with the azimuthal angle of the incidence plane, $\phi_{\inc}$, 
giving rise to the Airy spirals formed due to 
rotation of the polarizatiion ellipses. 

Note  that the above discussed case of normal incidence
%with $\theta_{\inc}=0$  
corresponds to the centre of images.
Referring to Fig.~\ref{fig:pattern}a-e
it can be seen that, for the central polarization ellipse,
counterclockwise rotation is the only effect caused by
the increase in the defect twist angle. 
Clearly, this effect is in agreement with the results
presented in Fig.~\ref{fig:cross_transm}b.
Fig.~\ref{fig:pattern}a-e show that,
as opposed to the central ellipse, 
the conoscopic images and the polarization resolved patterns
as a whole rotate clockwise 
with the defect angle, $\Delta\phi$.

In addition to the rotation effect,
there are defect mode sharp resonance dips
in the distribution of the transmission amplitude.
These dips 
form the dark thin rings (circles)
which are indicated by arrows in the conoscopic images~\ref{fig:pattern}b-d.
It can be seen that, at small defect angles,
the defect mode ring develops near the band gap edge
(see Fig.~\ref{fig:pattern}b)
and it shrinks to the point located at the origin 
when the defect angle increases reaching its 
limiting value $\Delta\phi=90$~deg.
(see Fig.~\ref{fig:pattern}d).
As is shown in Fig.~\ref{fig:pattern}e, the defect mode
resonances disappear at $\Delta\phi>90$~deg.
This effect is additionally illustrated in Fig.~\ref{fig:pattern}f
where the incidence angle $\theta_{\dd}$ giving the radius
of the defect mode ring
is plotted against the defect twist angle, $\Delta\phi$.
The radius is shown to be a monotonically decreasing function
of the defect angle that vanishes at $\Delta\phi=\pi/2$.

By contrast to the intensity conoscopic images,
the polarization resolved patterns appear to be insensitive
to the presence of the defect mode resonances.
So, in these patterns, we have the defect induced rotation 
as the only significant effect. 
Note that, for normally incident incoming light,
the analogous results on insensitivity of the polarization characteristics
immediately follow from the curves shown in 
Fig.~\ref{fig:cross_transm}b.

So,  we may conclude that,
in the conoscopic patterns of thick CLC cells with twist defects
at varying defect angle, the most important effects are:
(a)~the defect angle induced rotation of the pictures as a whole and 
(b)~the defect mode ring shrinking to the origin
with the defect twist angle.

It should be emphasized that, 
in  thin layers,
the patterns are expected to reveal more intricate behaviour
as their structure will be complicated by additional interference effects.
These results  along with the comprehensive
theoretical analysis omitted in this work
will be published elsewhere.

%\bibliographystyle{apsrev}
%\bibliographystyle{ol}
%\bibliography{optics,polymer,scatter,lc,quant,hk,flc,qft,math}

\end{document}